\documentclass[fleqn,10pt]{wlscirep}
\usepackage[utf8]{inputenc}
\usepackage[T1]{fontenc}
\title{Heat flow due to time-delayed feedback}

\author[1,*]{Sarah A.~M.~Loos}
\author[1]{Sabine H.~L.~Klapp}
\affil[1]{ Institut f\"ur Theoretische Physik,
  Hardenbergstr.~36,
  Technische Universit\"at Berlin,
  D-10623 Berlin,
  Germany}

\usepackage{changes}
\affil[*]{sarahloos@itp.tu-berlin.de}

\newcommand{\finalstate}[1][]{%
  \renewcommand{\deleted}[4][]{}
  }%

\begin{abstract}
Many stochastic systems in biology, physics and technology involve discrete time delays in the underlying equations of motion, stemming, e.\,g., from finite signal transmission times, or a time lag between signal detection and adaption of an apparatus. From a mathematical perspective, delayed systems represent a special class of non-Markovian processes with delta-peaked memory kernels. It is well established that delays can induce intriguing behaviour, such as spontaneous oscillations, or resonance phenomena resulting from the interplay between delay and noise.

However, the thermodynamics of delayed stochastic systems is still widely unexplored. This is especially true for continuous systems governed by nonlinear forces, which are omnipresent in realistic situations. We here present an analytical approach for the net steady-state heat rate in classical overdamped systems subject to time-delayed feedback. We show that the feedback inevitably leads to a finite heat flow even for vanishingly small delay times, {and detect the nontrivial interplay of noise and delay as the underlying reason.} 
{To illustrate this point, and to provide an understanding of the heat flow at small delay times below the velocity-relaxation timescale, we compare with the case of underdamped motion where the phenomenon of ``entropy pumping'' has already been established.}
 Application to an exemplary {(overdamped)} bistable system reveals that the feedback induces heating as well as cooling regimes and leads to a maximum of the medium entropy production at coherence resonance conditions. These observations are, in principle, measurable in experiments involving colloidal suspensions.
\end{abstract}
\begin{document}
\flushbottom
\maketitle
\thispagestyle{empty}

\section*{Introduction}
A finite heat flow is a generic feature of systems out of thermal equilibrium. 
In the last decades, special interest has been devoted to heat exchange and other thermodynamic properties of small (mesoscopic) systems coupled to a bath, which are noisy {\em per se}~\cite{Sekimoto2010, Seifert2012}. Stochastic thermodynamics (ST)
has emerged as an elegant and consistent framework to generalize thermodynamic notions to the level of noisy trajectories and to systems far from equilibrium \cite{Esposito2012}, with numerous applications to soft matter \cite{Speck2016}, biological\cite{Parrondo2015, Barato2015}, 
 and quantum systems\cite{Esposito2009,Strasberg2013}. 
Many fundamental concepts, however, are based on the Markov assumption, although in real-world systems memory effects are in fact often not negligible. While extensions towards several non-Markovian systems have been carried out in the past\cite{Speck2007,Vaikuntanathan2009,Kutvonen2015,Hasegawa2011,Garcia2012,Roche2015,Whitney2018,Mai2007,Schmidt2015,Cui2018}, the application to continuous systems with discrete \textit{time-delays} (i.\,e., delta-like memory kernels in the equations of motion) is still in its infancy.
 Delays can be of intrinsic nature as in neural systems~\cite{Longtin1990,Cabral2014}
and laser networks\cite{Schoell2016,Kane2005},
 or be generated externally, e.\,g., by a feedback protocol with time lag between signal detection and action of control\cite{Loos2014,Khadka2018,Mijalkov2016,Bruot2011}. 
 Moreover, since time-delay is known to enhance control strategies, it is often included intentionally, for example in Pyragas control\cite{Schoell2016}. Despite their major importance, delayed systems are still little understood from a thermodynamic perspective, especially in regard to the interplay of delay and nonlinearities.

Recent research\cite{Munakata2014,Rosinberg2015,Rosinberg2017,Munakata2009} has revealed that ST of delayed continuous systems is indeed quite involved, even in the absence of nonlinearities. For linear cases, it has been explicitly shown that in the long-time limit a non-equilibrium steady state (NESS) with finite entropy production\cite{Munakata2009} is approached (in the absence of time-dependent forces), as the delay pushes the system out of equilibrium. Nonequilibrium inequalities have been found\cite{Rosinberg2015} that generalize the second law and provide bounds to the extractable work. Furthermore, the fluctuations of work, heat and entropy production have been investigated\cite{Rosinberg2017}. 
One crucial issue in the thermodynamic description of delayed systems is the acausality of time-reversed processes appearing in the path integral representation of fluctuating heat
$q$ and entropy production~\cite{Munakata2014,Rosinberg2015,Rosinberg2017}. 
Contrary to the Markovian case~\cite{Seifert2012}, the total average entropy production $\Delta {S}_\mathrm{tot}$ of a delayed system in a NESS differs from the medium entropy $\Delta {S}_\mathrm{m}\mathcal{T}\!=\!\langle q \rangle_\mathrm{ss} $ (where $\mathcal{T}$ is the heat bath's temperature and $\langle .. \rangle_\mathrm{ss}$ denotes NESS ensemble averages). In particular, the second law does not impose nonnegativity on $\Delta {S}_\mathrm{m}$ alone\cite{Munakata2014,Rosinberg2015,Rosinberg2017}. 
However, while these statements are generic, explicit expressions for the thermodynamic quantities are, so far, only available for systems governed by linear forces~\cite{Munakata2009,Munakata2014,Rosinberg2015,Rosinberg2017}, 
thus excluding wide classes of physically interesting processes which exclusively arise in nonlinear systems. 

As a step in this direction, we here apply ST to investigate the \textit{heat rate} $\dot{Q}=\langle\delta q /\mathrm{d}t\rangle_\mathrm{ss}$ of a classical nonlinear delayed system. By considering $\dot{Q} = \dot{S}_\mathrm{m} \mathcal{T} $ rather than $\dot{S}_\mathrm{tot}$, we avoid the above-mentioned problem induced by acausality, and at the same time consider a key thermodynamic quantity and nontrivial part of the total entropy production which already provides important physical insight into the thermodynamics of delayed systems. This strategy enables us to address several fundamental questions: What is the impact of time-delayed feedback on heat exchange and entropy production? Has the overdamped limit consequences for the thermodynamic description? Do thermodynamic quantities reflect delay-{\em induced} dynamical behaviour? An example are spontaneous oscillations occurring in many {nonlinear} delayed systems {due to their infinite dimensionality}~\cite{Mackey1994,Schoell2008}.

To this end, we consider a simple exemplary system composed of an overdamped particle subject to nonlinear static forces and a deterministic (i.\,e., error-free), {\em continuous} linear feedback force with a discrete time delay $\tau \geq 0$. Such a force is imposed, e.\,g., by optical tweezers~\cite{Blickle2012, Kotar2010}. 
The idealized heat bath is assumed to remain at equilibrium, unaffected by the feedback. Our calculation explicitly predicts that delay alone induces a finite heat flow whose direction is tunable. This means, in particular, that the delayed force can generate a steady heat flow from the bath to the particle, i.\,e., feedback cooling. We moreover unravel an important {universal heat flow caused by the nontrivial interplay between delay and noise, which also occurs for underdamped motion. We discuss its physical origin in detail and propose that this heat flow is linked to the \textit{entropy pumping} known for underdamped systems with velocity-dependent feedback,  i.\,e., ``molecular refrigerators''\cite{Kim2004, Kim2007, Munakata2014}.} {We further detect discontinuous behaviour at $\tau\to 0$.} Discussing the application to a paradigmatic bistable system~\cite{Roldan2014}, we evaluate the heat rate via several approximations and by numerical simulations. In particular, we consider the medium entropy production near {\em coherence resonance}\cite{Schoell2008,Gang1993,Zakharova2013,Geffert2014} (CR), 
that is, the appearance of regular
positional oscillations {at a finite thermal energy} caused by the interplay of nonlinearity, noise, and delay~\cite{Tsimring2001,Masoller2002,Masoller2003,Xiao2016}. 
{Importantly, in contrast to stochastic resonance, there is no periodic external driving.}
 Combining the discretized (Master) equation approach of\cite{Tsimring2001} with ST, we show that the medium entropy production has a maximum at CR, and provide an {analytical} explanation for this maximum, which has, so far, only
been detected numerically~\cite{Xiao2016}.

%
%
%
\subsection*{Model}
We consider stochastic processes described by the overdamped Langevin equation (LE)~\cite{Loos2017}
\begin{align}\label{EQ:LE}
 \gamma{\mathrm{d} X (t)} =  F_\mathrm{con}[X(t)]{\mathrm{d}t} + F_\mathrm{d}[X(t-\tau)]  {\mathrm{d}t}+  \gamma \sqrt{2 D_0}\, \xi(t)\,{\mathrm{d}t}=  -\sum_{i=1}^{m} a_i X(t)^i{\mathrm{d}t} -b X(t-\tau)  {\mathrm{d}t}+   \gamma\sqrt{2 D_0}\, \xi(t)\,{\mathrm{d}t},
\end{align}
with $a_i, b \in \mathbb{R}$, {where the {total} deterministic force $F(x,x_\tau)=F_\mathrm{con}(x)-b x_\tau$ depends on the instantaneous, $X(t)$, and on the delayed particle position $X(t-\tau)$. {We assume that the conservative force can be expressed as a polynomial (which holds indeed for a wide class of nonlinear potentials), while the feedback control $F_\mathrm{d}$ is chosen to be linear corresponding, e.\,g., optical tweezers~\cite{Blickle2012}.} $\xi$ denotes Gaussian white noise with $\left\langle \xi(t)\right\rangle=0$ and $\left\langle \xi(t)\xi(t')\right\rangle=\delta (t-t')$, while $\gamma$ and $D_0$ are the friction and diffusion coefficients satisfying\cite{Risken1984} $\gamma D_0\!=\!k_{\mathrm{B}}{\cal T}$, with $k_{\mathrm{B}}$ being the Boltzmann constant. Due to the appearance of the delayed position in (\ref{EQ:LE}), the corresponding Fokker-Planck equation {(given at the end of this report in the Technical aspects section)} for the probability density function $\rho_{}(x,t)$ (PDF) is an infinite hierarchy~\cite{Guillouzic1999, Loos2017, Frank2005a, Frank2005}. We consider natural boundary conditions, i.\,e., $\rho_{} =$ $\partial_x \rho_{} \rightarrow 0$ for ${x\rightarrow \pm \infty}$, and focus on NESSs, where $\partial_t \rho_{}(x,t) =0$.
Following the ST framework of Sekimoto~\cite{Sekimoto2010,Munakata2009},
the fluctuating heat $\mathrm{\delta}q$ flowing to the reservoir during the infinitesimal time $\mathrm{d}t$, the increment of internal energy $\mathrm{d}u $, and the work $\mathrm{\delta}w$ done by the nonconservative (delayed) forces, are given by 
\begin{align}
\mathrm{\delta}q(t)&= \gamma\big[{\mathrm{d}  X (t)}/{\mathrm{d}t}-\sqrt{2D_0}\,\xi(t)\big]\circ \mathrm{d} X(t),\label{EQ:DefQ}\\
\mathrm{d}u(t)&= - F_\mathrm{con}\left[X(t)\right]\circ \mathrm{d}X(t), \label{EQ:DefE} \\
\mathrm{\delta}w(t)&=\mathrm{d}u(t)+\mathrm{\delta}q(t)=-b X(t\!-\!\tau)\circ \mathrm{d} X(t). \label{EQ:DefW}
\end{align}
The $\circ$-symbol indicates usage of Stratonovich calculus. Plugging the LE~(\ref{EQ:LE}) into Eqs.~(\ref{EQ:DefE},\ref{EQ:DefW}) results in the NESS ensemble averages 
\begin{align}
 \dot{U}\equiv\left\langle {\mathrm{d} u/\mathrm{d}t}\right\rangle_\mathrm{ss}=& \frac{1}{\gamma}\sum_{i=1}^{m} \Big \{ a_i \gamma \sqrt{2 D_0} \left\langle  X^i \xi\right\rangle_\mathrm{ss} - {a_i b} \,C_i(\tau) 
-{a^2_i} \left\langle X^{2i} \right\rangle_\mathrm{ss}-\sum_{j > i }^{m} {2 a_ia_j} \left\langle X^{i+j} \right\rangle_\mathrm{ss} \Big\},\label{EQ:E2} \\
 \dot{W}\equiv\left\langle{\delta w/\mathrm{d}t}\right\rangle_\mathrm{ss}=& \frac{b}{\gamma} \Big\{ \sum_{i=1}^{m}a_i C_i(\tau)+{b}\left\langle  X^2\right\rangle_\mathrm{ss} 
-{\gamma \sqrt{2D_0}\left\langle  X(t-\tau) \xi(t)\right\rangle_\mathrm{ss}}\Big\}.\label{EQ:W2}
\end{align}
Due to the time delay, the thermodynamic quantities depend on the spatial autocorrelation functions at time difference $\tau$, $C_i(\tau)=\left\langle  X(t)^i  X(t-\tau)\right\rangle_\mathrm{ss}$, mirroring the non-Markovian nature of~(\ref{EQ:LE}). 
\section*{Heat rate and medium entropy production}
For nonlinear systems, the linear response function method used in\cite{Munakata2009}, which bases on the Laplace-transformation, cannot be applied. In the following, we present an alternative approach for the heat rate. In particular, we derive exact expressions for~(\ref{EQ:E2},\ref{EQ:W2}) which only involve positional moments at one time. To this end, we utilize two relations that can be derived by projecting the Fokker-Planck equation onto the positional moments and subsequently inserting the LE~\cite{Frank2001}.

First, the $C_i(\tau)$-terms can be substituted by the relation~\cite{Frank2001} (valid $\forall n\geq 1$)
\begin{align}\label{EQ:Moments-FPE}
{b} \,C_n(\tau) = {\gamma} D_0 \,n  \left\langle  X^{n-1} \right\rangle_\mathrm{ss} 
- \!\!\sum_{i=1}^{m} a_i \left\langle  X^{n+i} \right\rangle_\mathrm{ss}.
\end{align}  
Second, the instantaneous noise--position cross correlations in~(\ref{EQ:E2}) can be replaced by positional moments via~\cite{Frank2001}
\begin{equation}\label{EQ:Moments-FPE-LE}
{ \left\langle  X(t)^{n} \xi(t) \right\rangle= n \,\sqrt{D_0/2}  \, \left\langle  X(t)^{n-1} \right\rangle_\mathrm{ }},~\forall n\geq 1.
\end{equation}
{A derivation of Eqs.~(\ref{EQ:Moments-FPE},\ref{EQ:Moments-FPE-LE}) can be found at the end of this report in the \textit{Technical aspects} section.}
Finally, we combine Eq.~(\ref{EQ:Moments-FPE-LE}) with a causality argument to evaluate $ \left\langle  X(t-\tau) \xi(t) \right\rangle$ at $\tau \neq 0$. 
Because no physical quantity can be influenced by future noise, statistical independence follows and $\left\langle  X(t) \xi(t')\right\rangle =\left\langle  X(t)\right\rangle \left\langle\xi(t')\right\rangle\equiv 0$ must hold $\forall t'>t$. Hence
\begin{equation}\label{EQ:limitXGamma}
\left\langle  X(t-\tau) \xi(t)\right\rangle = \sqrt{\frac{D_0}{2}} \delta_\tau=
\begin{cases}
\sqrt{{D_0}/{2}},& \tau=0\\
0, &\tau > 0
\end{cases}.
\end{equation}
Substituting Eqs.~(\ref{EQ:Moments-FPE}) and~(\ref{EQ:Moments-FPE-LE}) at $n=1,2,..,m$, respectively, into Eq.~(\ref{EQ:E2}) 
yields $\left\langle \mathrm{d} u/\mathrm{d}t\right\rangle_\mathrm{ss}\!=\!0$. This is expected, since the net impact of conservative forces should vanish in a NESS~\cite{Seifert2012}. Thus $\dot{W} \!=\!\left\langle{\delta q/\mathrm{d}t}\right\rangle_\mathrm{ss}\equiv \dot{Q}$. Substituting~(\ref{EQ:Moments-FPE}) at $n=1,2,..,m$ into Eq.~(\ref{EQ:W2}) {(please note that this simplification step is possible due to the polynomial form of force $F_\mathrm{con}$)} and using~(\ref{EQ:limitXGamma}) further yields 
\begin{align}\label{EQ:W3}
\dot{Q}= & ~\dot{W} = \sum_{i=1}^{m}\!\! \Big\{i a_i    D_0 \left\langle  X^{i-1} \right\rangle_\mathrm{ss} 
- \sum_{j=1}^{m}\frac{a_i{a_j}}{\gamma} \left\langle  X^{i+j} \right\rangle_\mathrm{ss} \Big\} 
+ \frac{b^2}{\gamma}\left\langle  X^2\right\rangle_\mathrm{ss}
-{b}  D_0 \delta_{\tau}  
=\dot{S}_\mathrm{m}\frac{\gamma D_0 }{k_\mathrm{B}} .
\end{align}

Equation~(\ref{EQ:W3}) is an exact expression only involving one-time ensemble averages over $X^n$. $\dot{Q}$ and $\dot{S}_\mathrm{m}$ can therefore be computed directly from the steady-state {one-time PDF}, and hence, on the basis of several approximations known from the literature. {The $\delta$-term suggests discontinuous behaviour of $\dot{Q}$ at $\tau \to 0$. But in order to study this limit properly, one also has to investigate the behaviour of the PDF to clarify whether the moments behave continuously. For nonlinear systems, this is a nontrivial task on its own, as, in fact, no exact solutions for the one-time PDF in the presence of delay are known. {However, we can nevertheless address this question analytically, since}} the approximative PDFs become \textit{exact} in the Markovian limits~\cite{Guillouzic1999,Loos2017}, rendering exact results from Eq.~(\ref{EQ:W3}).

A detailed description of the approximation schemes is beyond the scope of this report. We refer the interested reader to\cite{Loos2017} and references therein, and here only review some main aspects in brief. The small\,$\tau$ approximation~\cite{Guillouzic1999} is obtained by a first-order Taylor expansion around $\tau=0$ in the LE, effectively rendering a Markovian system (with exponentially decaying correlations). Two other schemes start from the Fokker-Planck equation for the one-time PDF, which involves the two-time PDF~\cite{Guillouzic1999} $\rho_2(x,t;x',t-\tau)$ {(see \textit{Technical aspects} section)}. The force-linearisation closure~\cite{Loos2017} (FLC) approximates $\rho_2$ by the one from the corresponding linear delayed system. The perturbation theory\cite{Frank2005a, Frank2005} (PT) approximates it with $\rho_2$ of the corresponding Markovian process without delay force. For a doublewell potential, the PT requires an additional approximation, since the corresponding two-time PDF is not known. As in~\cite{Loos2017}, we use the small-time propagator (PT-st), or the Ornstein-Uhlenbeck (PT-OU) approximation.   
{In the following, we will investigate the heat rate in the two (Markovian) limits, where the approximations become exact.}
\subsection*{{Limit of vanishing feedback strength}}
{In the limits where Markovianity is recovered, one expects that the system equilibrates (because of the absence of a driving force), and therefore $\dot{Q}=\dot{S}_\mathrm{m}=0$.}
The first limit corresponds to $b\rightarrow 0$, i.\,e., vanishing feedback strength.
By construction, the one-time steady-state PDFs from the FLC and from the PT become exact at $b_{} \rightarrow 0$. They converge in a {\em continuous} manner to the equilibrium (Boltzmann) distribution $\rho^{b=0}_\mathrm{ss }$, as can be easily seen from the PT-st result~\cite{Loos2017} $\rho^\mathrm{PT}_\mathrm{ss}$:
\begin{align}
\rho^\mathrm{PT}_\mathrm{ss }(x)  =Z^{ } \exp{ \left \{ -\frac{ \left[1- (b_{} \tau/\gamma) \right] \left[V_\mathrm{s}(x) +( {b_{}}/{2}) x^2\right] }{\gamma D_0} \right\}} 
~\stackrel{b\to 0}{\longrightarrow}~ \rho^{b=0}_\mathrm{ss }(x)=Z^{ }\exp{ \left [  -\frac{V_\mathrm{s}(x) }{\gamma D_0} \right ]},
\end{align}
with $V'_\mathrm{s}(x)=-F_\mathrm{con}(x)=\sum_{i=1}^{m} a_i x^i$ and normalization constant $Z$. After performing a partial integration step and plugging in the natural boundary conditions, the $(i-1)^{st}$ positional moment at $\tau=0$ can therewith be expressed as
\begin{align}\label{EQ:BetaLim}
\left\langle X^{i-1} \right\rangle_\mathrm{ss} =& \int_{-\infty}^{\infty} Z^{ } \exp{\left[-\frac{V_\mathrm{s}(x) }{\gamma D_0}\right]} x^{i-1}  \mathrm{d}x
=\left[\rho^{b=0}_\mathrm{ss }(x) \,\frac{ x^{i} }{i} \right ]_{-\infty}^{\infty} 
- \int_{-\infty}^{\infty} \frac{F_\mathrm{con}(x)}{\gamma D_0}Z^{ } \exp{\left[-\frac{V_\mathrm{s}(x) }{\gamma D_0}\right]} \frac{ x^{i}}{i } \mathrm{d}x  
\nonumber \\
=&   \sum_{j=1}^{m} \frac{a_j}{i \gamma D_0} \int_{-\infty}^{\infty} x^j \rho^{b=0}_\mathrm{ss }(x)   x^{i} \mathrm{d} x 
= \sum_{j=1}^{m}   \frac{a_j}{i \gamma D_0} \left\langle X^{j+i} \right\rangle_\mathrm{ss}, ~~\forall i\geq 1 
\\
\Rightarrow \sum_{i=1}^{m} i \,a_i  D_0 \left\langle X^{i-1} \right\rangle_\mathrm{ss} =&  \sum_{i=1}^{m}\sum_{j=1}^{m}   \frac{a_ja_i}{\gamma } \left\langle X^{j+i} \right\rangle_\mathrm{ss}.
\end{align}
Plugging the last expression into Eq.~(\ref{EQ:W3}), we finally conclude that $\lim_{b_{} \rightarrow 0}\dot{S}_\mathrm{m} = 0$ in a \textit{continuous} manner.
\subsection*{{Limit of vanishing delay time}}
The other relevant limit is that of vanishing delay time, $\tau\rightarrow 0$. 
The one-time PDF~\cite{Loos2017} from the small $\tau$ expansion~\cite{Guillouzic1999} 
 \begin{equation}\label{EQ:Veff-smallTau}
 \rho^{s\tau}_\mathrm{ss }(x)=Z^{ }\exp{\left [ -\frac{V_\mathrm{s}(x)+ (b_{}/2) x^2}{(\gamma + b_{}\tau) D_0} \right ]}
 ~\stackrel{\tau\to 0}{\longrightarrow}~ \rho^{\tau=0}_\mathrm{ss }(x)=Z^{ } \exp{\left [ -\frac{V_\mathrm{s}(x) + (b_{}/2) x^2 }{\gamma D_0} \right ]}
 \end{equation}
clearly becomes exact {for $\tau \rightarrow 0$}, where it {\em continuously} converges to the Boltzmann distribution $\rho^{\tau=0}_\mathrm{ss }$.
 Performing steps analogously to Eq.~(\ref{EQ:BetaLim}), we obtain
 \begin{align}\label{EQ:TauLim1}
 \left\langle X^{i-1} \right\rangle_\mathrm{ss} 
 = \left[Z^{ } \rho^{\tau=0}_\mathrm{ss }(x) \, ({x^{i}}/{i}) \right]_{-\infty}^{\infty}- \int_{-\infty}^{\infty}  \frac{F_\mathrm{con}(x)- b_{} x }{i \gamma D_0}x^{i}\rho^{\tau=0}_\mathrm{ss }(x) \mathrm{d}x 
 = \sum_{j=1}^{m}   \frac{a_j }{i \gamma D_0} \left\langle X^{j+i} \right\rangle_\mathrm{ss} +  \frac{b_{}}{i \gamma D_0}  \left\langle X^{i+1} \right\rangle_\mathrm{ss} ,
 \end{align}
 for all $ i\geq 1$ at $\tau\rightarrow 0$, which results in the identities
 \begin{align}
 \sum_{i=1}^{m} i \,a_i  D_0 \left\langle X^{i-1} \right\rangle_\mathrm{ss} = & \sum_{i=1}^{m}\sum_{j=1}^{m}  \frac{a_ja_i}{\gamma} \left\langle X^{j+i} \right\rangle_\mathrm{ss} +\sum_{i=1}^{m}  \frac{b_{} a_i}{\gamma}  \left\langle X^{i+1} \right\rangle_\mathrm{ss},~~~~~ 
  \left\langle X^{0} \right\rangle_\mathrm{ss} \equiv 
  1=   \frac{b_{}}{\gamma D_0}  \left\langle X^{2} \right\rangle_\mathrm{ss}  + \sum_{j=1}^{m}   \frac{ a_j }{\gamma D_0} \left\langle X^{j+1} \right\rangle_\mathrm{ss}.
 \end{align}
 Plugging both into Eq.~(\ref{EQ:W3}) in an iterative manner, yields the \textit{discontinuous} limit
\begin{align}\label{EQ:discont}
\lim_{\tau \rightarrow 0}\dot{S}_\mathrm{m}=\lim_{\tau \rightarrow 0} \frac{k_\mathrm{B}{b}}{\gamma }(
 1
- \delta_{\tau}  )  =
\begin{cases}
0, &\tau = 0\\
{k_\mathrm{B}{b}}/{\gamma } ,& \tau>0 
\end{cases}.
\end{align}
This is our first main result.
This ubiquitous jump-discontinuity at $\tau\rightarrow 0$ (see Fig.~\ref{FIG1} for an example) indicates an abrupt qualitative change of the thermodynamics when non-Markovianity sets in. It arises due to the discontinuity of the noise--position cross correlations~(\ref{EQ:limitXGamma}) at the onset of causal relationship. 
Remarkably, the apparent offset of ${k_\mathrm{B}{b_{}}}/{\gamma }$ is independent of the details of the potential landscape. We note that the apparent offset has already been observed and discussed in the context of linear systems~\cite{Munakata2014,Rosinberg2015,Rosinberg2017, Munakata2009}. In~\cite{Munakata2009} it has been considered as a consequence of {inconsistent} usage of Ito {and Stratonovich} calculus. However, as shown here, it is not a mathematical error but also arises within consistently applied Stratonovich calculus. The underlying reason is {the interplay of white noise and delay below} the short (ballistic) relaxation timescale~\cite{Gardiner2002}, which becomes relevant as $\tau \rightarrow 0$. 
{To further elucidate the behaviour at $\tau\to 0$, we also briefly consider the limit in the \textit{underdamped} case, where the LE
\begin{align}\label{EQ:LE-underdamped}
m \ddot{X}(t)  =- \gamma{\dot{X} (t)}+  F_\mathrm{con}[X(t)] -b X(t-\tau)  +  \gamma\sqrt{2 D_0}\, \xi(t) \text{~~(underdamped LE)},
\end{align}
 additionally involves an inertial term with mass $m$, yielding ballistic motion below the velocity-relaxation timescale, $\gamma/m$. For this system, the heat can, in principle, be calculated using the same expression as in the overdamped case, see Eq.~(\ref{EQ:DefQ},\ref{EQ:DefW})
As has been shown previously\cite{Munakata2014} for linear systems, $\dot{Q}$ smoothly decays to zero with $\tau$. However, the response function method used in\cite{Munakata2014} is only applicable to linear systems. For a better comparison, we here consider the correlation $\left\langle {X}(t-\tau) \xi(t) \right\rangle$ appearing in Eq.~(\ref{EQ:W2}), which, in our framework, ``causes'' the discontinuity from a mathematical point of view [compare Eq.~(\ref{EQ:limitXGamma},\ref{EQ:discont})]. As opposed to the overdamped case, this correlation is indeed continuous at $\tau=0$,
\begin{align}\label{EQ:Contin-x-dx-Corr-underdamped}
\left\langle {X}(t-\tau) \xi(t) \right\rangle =&\, 0, ~ \forall \tau \geq 0
\end{align}
(not only for linear systems), indicating that $\dot{Q}$ behaves continuous. The derivation of (\ref{EQ:Contin-x-dx-Corr-underdamped}) is given in the \textit{Technical aspects }section.
}
\\
{Turning back to the overdamped system, apart from the discontinuous limit, we find from Eq.~(\ref{EQ:discont}) that }for $\tau>0$, $\dot{S}_\mathrm{m}$ and $\dot{Q}$ are nonzero, proving the true non-equilibrium nature of this steady state. {According to (\ref{EQ:discont}), }the rates have negative values at small $\tau$, if $b<0$, which is always the case in an optical tweezers setup (see below). The negative signs indicate a steady heat flow from the bath to the particle (feedback cooling). This is a delay-induced phenomenon, which would be impossible in the Markovian counterpart of this system due to the second law\cite{Kim2007}. The fact that here $\dot{S}_\mathrm{m}\!<\!0$ (the bath constantly loses entropy) underlines that further entropic terms must contribute to the (nonnegative) total entropy production (as discussed for underdamped dynamics in~\cite{Munakata2014,Rosinberg2015,Rosinberg2017}). In the cooling regime, the amount of medium entropy loss thus provides a lower bound to the entropic cost \cite{Munakata2014,Rosinberg2015,Rosinberg2017} of the feedback.}
{
Analytical evaluation of the heat rate for (linear and nonlinear) example systems, which we will provide below, reveals that the (negative) heat flow is actually quite pronounced in the regime of small $\tau$ (compare with Fig.~\ref{FIG1}). Notably, this is also true for underdamped linear systems\cite{Munakata2014} (where $\dot{Q}$ only decays for $\tau$ below the velocity-relaxation timescale, compare with Fig.~1 in\cite{Munakata2014}). Before proceeding with the concrete examples, we here propose an explanation of this phenomenon.}
\subsubsection*{{Discussion of the behaviour for small delay times}}
{
When $\tau$ is so small that ${X}(t-\tau)\approx  {X}(t)- \tau \dot{X}(t-\tau)$, the feedback force $F_\mathrm{d}=-b X(t-\tau)$ gets a contribution proportional to the strongly fluctuating velocity. This changes the induced steady-state heat flow $\dot{Q}=\langle F_\mathrm{d}\circ \dot{X}(t) \rangle$ significantly. For $b<0$, $F_\mathrm{d}\sim  \tau b \dot{X}(t-\tau)$ is a \textit{friction-like} force. As is well-known for underdamped systems with velocity-dependent (delayed\cite{Munakata2014} or non-delayed\cite{Kim2004,Kim2007}) feedback, such a control amounts to medium entropy reduction due to \textit{entropy pumping}. {(Please note that a non-delayed velocity-dependent feedback $F_\mathrm{d}\sim b \dot{X}(t)$ drives a system out of equilibrium\cite{Kim2004,Kim2007}, contrary to a non-delayed position-dependent control)}. The reason is that the additional friction-like force reduces the thermal fluctuations of the particle, inducing an energy transfer from the bath to the particle, hence, a heat flow. (Reversely, the particle fluctuations are enhanced for $b>0$, yielding a positive heat flow.) This means, an additional ``entropy pumping'' contribution to the heat flow arises, explaining the enlarged heat flow at small delay times. Please note that this effect is independent of the question whether the overdamped limit is used, or not, and should be measurable in experimental setups.
\\
When $\tau$ gets below the velocity-relaxation time ($m/\gamma$), both system behave differently. For underdamped dynamics, the velocities are correlated, implying that the delay-induced heat flow $\dot{Q} \sim   \tau\,b \langle  \dot{X}(t-\tau) \dot{X}(t) \rangle \sim \tau\,e^{-\gamma \tau/m} $ eventually decays smoothly to zero (resulting in a maximum of $|\dot{Q}|$ around $\tau=m/\gamma$, see Fig.~1 in \cite{Munakata2014}). For overdamped dynamics, the velocities are, in contrast, uncorrelated (because the white noise directly acts on it $\dot{X}\sim \xi$), giving rise to the nontrivial limit $\tau\, \delta(\tau)$ of type ``$0 \times \infty$''. This yields a finite value as $\tau \to 0$, as we know from Eq.~(\ref{EQ:discont}). Hence, in an experimental setup, a position-dependent feedback is expected to induce heat flow, unless the (typically unavoidable) delay is well-below the velocity-relaxation timescale.} 
\subsection*{Linear delayed system}
\begin{figure}
\includegraphics[width=1\linewidth]{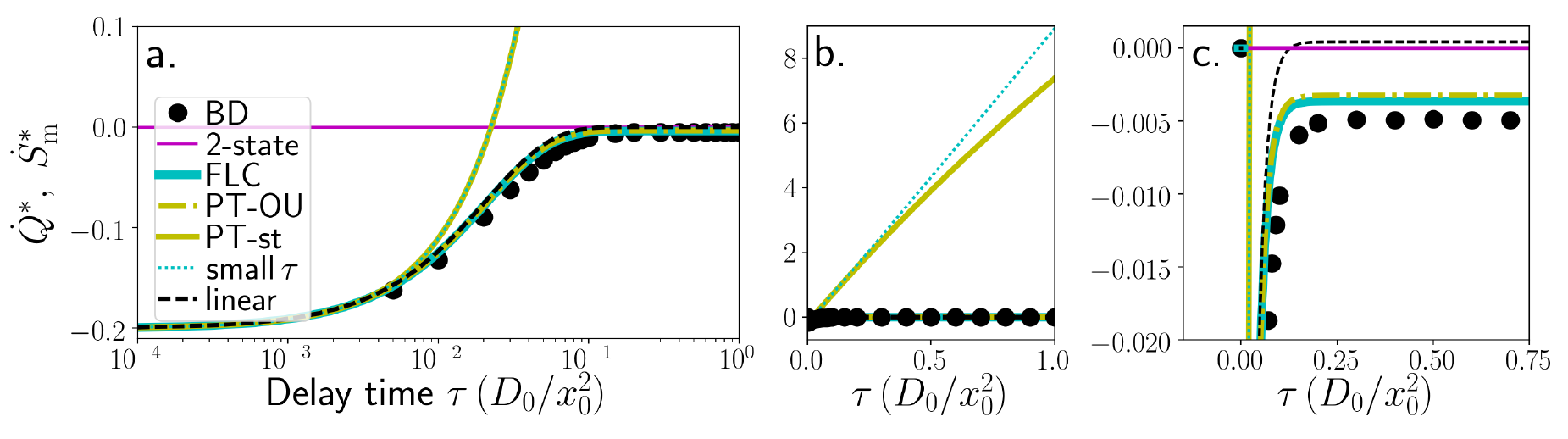}
\caption{Scaled NESS mean heat rate $\dot{Q}^* \!=\! \dot{Q}\,{x_0^2}/({\gamma D_0^2})$ and medium entropy production rate $\dot{S} _\mathrm{m}^* \!=\! \dot{S} _\mathrm{m}\,{x_0^2 }/({k_\mathrm{B} D_0 })$ vs. delay time, in the delayed bistable system at $V_0 = 6\gamma D_0$, $k = 0.2\gamma D_0$ and in the corresponding linearised system, obtained from numerics (BD) and from several approximations. a. logarithmic x-axis, b.+c. linear plots: b. zoom out, c. magnification.  
}\label{FIG1}
\end{figure}

For linear systems, where the force has the form $F =- a_1 x - b x_\tau$, the exact NESS {PDFs} are known~\cite{Kuechler1992}, and~(\ref{EQ:W3}) simplifies to the exact and closed expression
\begin{equation}\label{EQ:W4}
\frac{\dot{Q}}{ D_0}=\frac{\gamma \dot{S}_\mathrm{m}}{ k_\mathrm{B}}
= a_1   -{b \delta_{\tau} }  - ({b^2-a_1^2})\left[ 1+b \sinh\left(\sqrt{a_1^2-b^2}\,\tau /\gamma\right)\Big/{\sqrt{|a_1^2-b^2|}}\right]\left[a_1+b \cosh\left(\sqrt{a_1^2-b^2}\,\tau / \gamma\right)\right]^{-1} .
\end{equation} 
A plot of $\dot{Q}$ as a function of $\tau$ is given in Fig.~\ref{FIG1}. As expected, $\dot{Q}$ has an apparent offset at $\tau=0$ and $|\dot{Q}|$  \textit{grows} while $\tau$ decays to zero (until a saturation value is reached).
Equation~(\ref{EQ:W4}) is consistent with~\cite{Munakata2009} (where only the linear case is considered and a different approach is used), apart from the \textit{additional} $\delta_{\tau}$\textit{-term}. We stress that, consequentially, only Eq.~(\ref{EQ:W4}) correctly predicts $\dot{S}_\mathrm{m}(\tau\rightarrow 0)=0$. 
%
%
%
%
%
\subsection*{Application to a delayed bistable system}
We now consider a bistable system composed of a doublewell potential $V_\mathrm{s}\!=\! V_\mathrm{0} [ {(x/x_0)}^4 -2  {(x/x_0)}^2 ]$ with minima at $\pm x_0$ and potential barrier height $V_0$, supplemented by the delayed (optical tweezers~\cite{Blickle2012}) potential $V_\mathrm{d}=(k/2)[ {(x/x_0)}-{(x_\tau/x_0)}]^2$ with $0\!<\!k\!<\!4V_0$. The potentials yield a polynomial force $F=-\mathrm{d}(V_\mathrm{s}+V_\mathrm{d})/\mathrm{d}x=- a_3 x^3-a_1 x  -b x_\tau$ with coefficients $b=-kx_0^{-2}$, $a_{1}\!=\!(-4V_\mathrm{0} + k)x_0^{-2} $, $a_{3}\!=\! 4 V_\mathrm{0}x_0^{-4}$. This is a prototypical nonlinear noisy system which exhibits both: a nontrivial intrawell dynamics within the asymmetric potential wells, and a noise- and delay-induced dynamical state, i.\,e., positional oscillations between the wells~\cite{Masoller2002,Masoller2003, Tsimring2001, Xiao2016}. For this system, Eq.~(\ref{EQ:W3}) involves the even moments up to $6^{th}$ order. 
Performing Brownian dynamics (BD) simulations (see \textit{Technical aspects} section for details) to compute the NESS heat rate, both, from Eq.~(\ref{EQ:DefQ}) and~(\ref{EQ:W3}), we indeed find perfect agreement. In the following, we compare the BD results to those obtained from our analytical approach, i.\,e., Eq.~(\ref{EQ:W3}) combined with established approximations. Since the approximations are known to perform best when the particle is likely to stay around a potential minimum~\cite{Loos2017}, this approach seems appropriate in the low thermal energy regime ($\gamma D_0 \!\ll\! V_0$). At the end, we will introduce a complementary approach for larger noise levels. 
\subsubsection*{Low thermal energy -- intrawell dynamics}
Figure~\ref{FIG1} shows $\dot{Q}=\dot{W}$ and $\dot{S}_\mathrm{m}$, as functions of the delay time $\tau$, for an exemplary parameter setting in the low noise regime. 
The simulation results confirm the discontinuous $\tau$-limit and apparent offset as given by Eq.~(\ref{EQ:discont}), and the predicted feedback cooling. The PT-OU predictions are quantitatively very similar to the ones obtained by the FLC, while the PT-st yields very similar results to the small\,$\tau$ expansion. We will therefore mainly discuss the FLC and the small\,$\tau$ expansion in the following.
 
While the small $\tau$ expansion fails outside the Markovian limit, our analytical approach with FLC makes quantitatively correct predictions for all delay times considered in Fig.~\ref{FIG1}. The thermodynamic quantities grow with $\tau$, until they approach constant values. Interestingly, the saturation occurs at about $\tau\approx x_0^2/(2 D_0)$, i.\,e., on the timescale where the mean-squared displacement $2 D_0 t$ of a freely diffusing particle is in the range of the extent of a potential well $\approx x_0$ (i.\,e., the particle has explored the whole well within $\tau$). $x_0^2/(2 D_0)$ is thus an estimate of the intrawell relaxation time $t_\mathrm{rel}$ in the delayed system.
Figure~\ref{FIG1} also displays the results from the corresponding linear system, $F =- a_1 x - b_{} x_\tau$ [with $a_1 = -(8  V_\mathrm{0} + k)/x_0 $ and $b_{} = -k/x_0 $, as given by a second order Taylor expansion of $V_\mathrm{s}+V_\mathrm{d}$ around a (deterministically) stable fix point, i.\,e., $(x,x_\tau) = \pm(x_0,x_0)$]. Interestingly, $\dot{Q}$ is not only equivalent to the nonlinear case for very small $\tau$ [as expected from~(\ref{EQ:discont})], but also saturates on the same timescale.

\begin{figure}
\includegraphics[width=.53\linewidth]{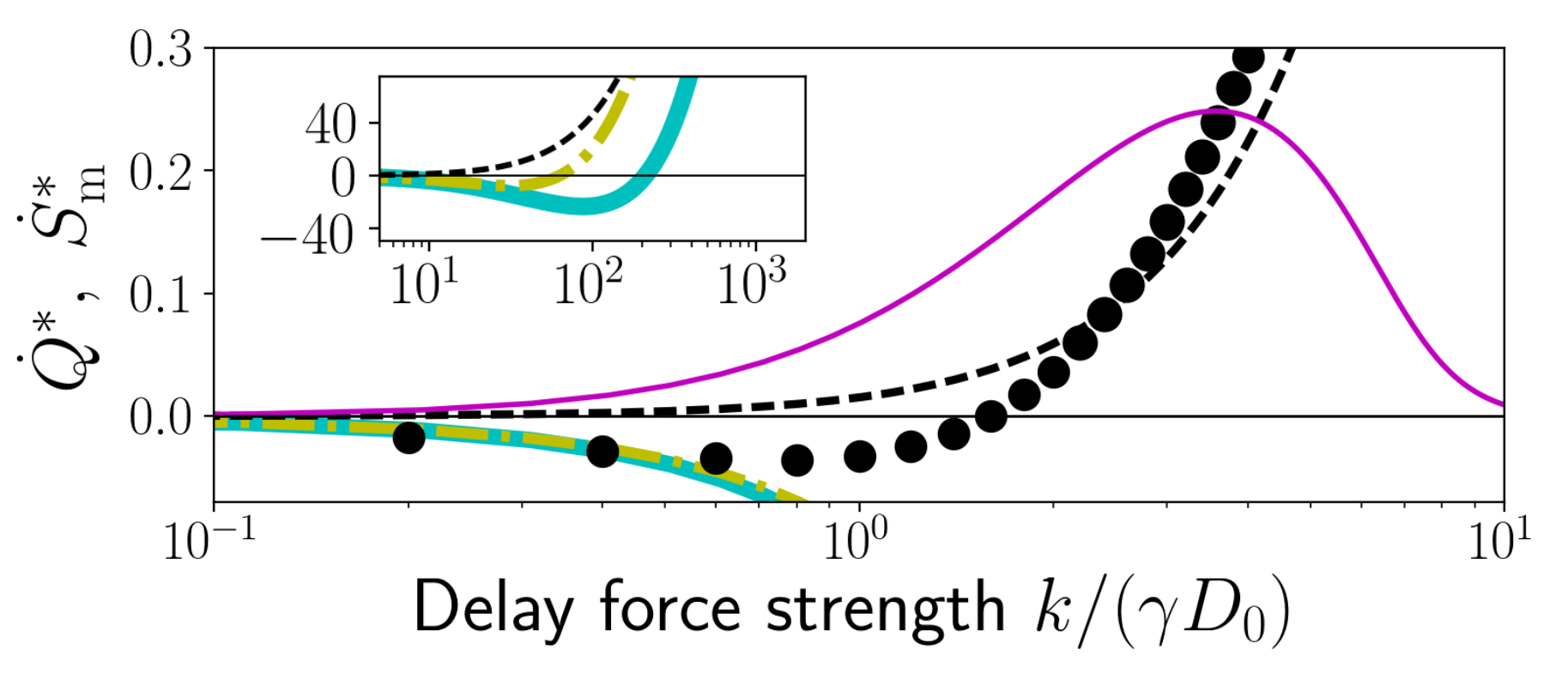}
\caption{Scaled NESS mean heat rate $\dot{Q}^* \!=\! \dot{Q}\,{x_0^2}/({\gamma D_0^2})$ and medium entropy production rate $\dot{S} _\mathrm{m}^* \!=\! \dot{S} _\mathrm{m}\,{x_0^2 }/({k_\mathrm{B} D_0 })$ vs. delay force strength $k$ over thermal energy $\gamma D_0$; at $V_0 = 4 \,\gamma D_0$, $\tau= x_0^2/D_0 $. Inset: zoom out. Colour code as in Fig.~\ref{FIG1}.}\label{FIG2}
\end{figure}

The agreement between BD and FLC persists at larger barrier heights $V_0$ or smaller $k$. 
However, for larger $k/(\gamma D_0)$, also the FLC approach breaks down, as can be seen in Fig.~\ref{FIG2}. The inset shows that the FLC nevertheless captures the qualitative behaviour. Upon increase of $k$ (i.\,e., the laser intensity in case of optical tweezers \cite{Blickle2012}), one may switch from feedback cooling to heating in the nonlinear case.
%
%
%
%
%
%
\subsubsection*{High thermal energy -- interwell dynamics}
When the thermal energy is sufficiently high, such that jump processes between the potential wells dominate the dynamics. 
the discussed approximations all break down by construction. The reason is that it is then the ``memory of a jump'' expressed in the non-Markovian $\rho_2$, which dominates the process and, e.\,g., triggers subsequent jumps, and precisely $\rho_2$ is what is (over)simplified in all approaches\cite{Loos2017}. In particular, the FLC effectively assumes an harmonic potential well out of which no jumps can occur, while the PT and small\,$\tau$ expansion render Markovian systems which don't exhibit spontaneous oscillations, as the latter are a delay-induced phenomenon.
This includes situations, where the interplay of noise and delay leads to spontaneous oscillations of $X$. However, we can still treat this regime via an alternative strategy, which bases on the discretised approach for multistable systems proposed by Tsimring and Pikovsky~\cite{Tsimring2001}. To this end, the $X$-dynamics is reduced to switching processes between \textit{two discrete} states $s=\pm x_0$ (corresponding to the two potential wells). There are two different transition rates corresponding to the two possible scenarios that the delayed and instantaneous state have the same, or opposite sign, respectively. These transition rates are approximated by the Kramers formula~\cite{ Loos2017, Tsimring2001, Kramers1940, Hanggi1990} within a quasistatic approximation. The latter is justified when the delay time is small compared to the intrawell relaxation time, $\tau\!\gg \!t_\mathrm{rel}$, and $\gamma D_0 < V_0$. {Further details including the calculation of the transition rates are given at the end of this report in the \textit{Technical aspects} section.}
 
\begin{figure}
\includegraphics[width=1\linewidth]{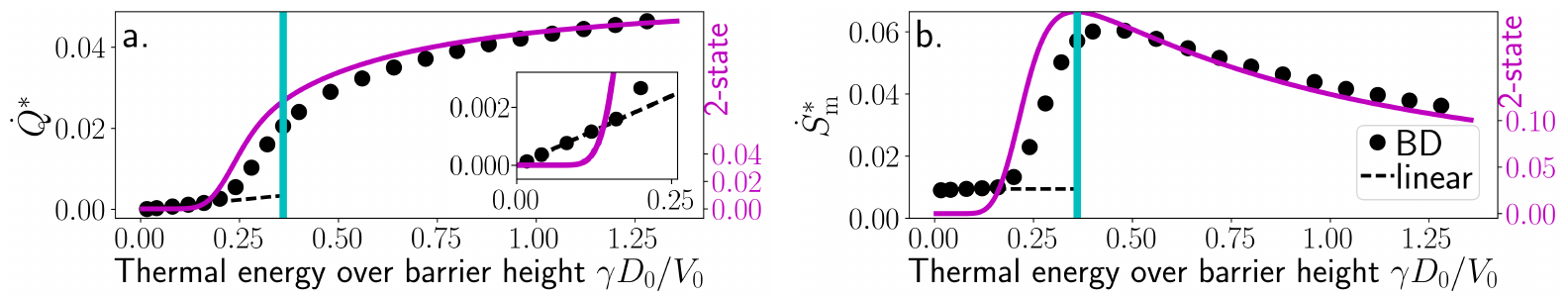}
\caption{ 
Analysis of the dependency on the thermal energy $\gamma D_0$ scaled with the barrier height $V_0$, in the delayed bistable system at $k=0.4\,V_0$, $\tau= 25\, \gamma x_0^2/V_0 $ and in corresponding linearised system. a. Scaled NESS heat rate $\dot{Q}^* \!=\! \dot{Q}\,({\gamma x_0^2}/{V_0^2})$ (the inset gives a magnification), b. medium entropy production rate $\dot{S} _\mathrm{m}^* \!=\! \dot{S} _\mathrm{m}\,\,({\gamma x_0^2}/{V_0} {k_\mathrm{B}})$.  
Magenta lines and y-axes: 2-state model. Vertical {cyan} lines: maximum of CR order parameter (see Fig.\,\ref{FIG4}).
}\label{FIG3}
\end{figure}

{The simplification of Eq.~(\ref{EQ:W2}) due to the 2-state reduction is twofold. First, the spatial correlation function $C_1(\Delta t)= \left\langle X^{i}(t)X(t-\Delta t) \right\rangle_\mathrm{ss}$ can now be calculated analytically~\cite{Tsimring2001}. Second, all other spatial autocorrelation functions are significantly simplified, since the intrawell-dynamics are omitted, yielding for $\tau>0$
\begin{align}
\left\langle X(t)^2 \right\rangle &\rightarrow \left\langle s(t)s(t) \right\rangle = \left\langle x_0^2 \right\rangle = x_0^2\\
C_i(\tau)= \left\langle X^{i}(t)X(t-\tau) \right\rangle&\rightarrow \left\langle s^{i}(t)s(t-\tau) \right\rangle = \begin{cases}
 x_0^{i} \left\langle s(t-\tau) \right\rangle = 0,& \forall \text{~even~} i\\
 x_0^{i-1}\left\langle s(t)s(t-\tau) \right\rangle = C_1(\tau)\, x_0^{i-1},& \forall\text{~odd~} i.
 \end{cases}
\end{align}
[Please note that the correlation $ \left\langle X(t-\tau) \xi(t) \right\rangle \rightarrow 0$ at $\tau>0$  due to causality.] Consequentially, Eq.~(\ref{EQ:W2}) simplifies to}
\begin{align}\label{EQ:Heat_2state}
\dot{W}=& (b/\gamma) \left( a_1+x_0^2 a_3 \right)\, C_1(\tau) + \left( b^2/\gamma \right)\, x_0^2 
= {k^2}/\left({\gamma}{x_0^2}\right) \left[ 1- C_1(\tau)/x_0^2 \right]\stackrel{!}{=} \dot{Q} = \dot{S}_\mathrm{m} \gamma D_0  / k_\mathrm{B},
\end{align}
where we have used that the mean change of internal energy must vanish in the NESS. Using the explicit formula for $C_1(\tau)$ {[see Eq.~(\ref{EQ:C1_Tsimring})]}, Eq.~(\ref{EQ:Heat_2state}) can readily be evaluated. The results are plotted in Fig.~\ref{FIG3}.
%
%
%
%
%
%
%
%
%
%
%
%

\begin{figure}
\includegraphics[width=1\linewidth]{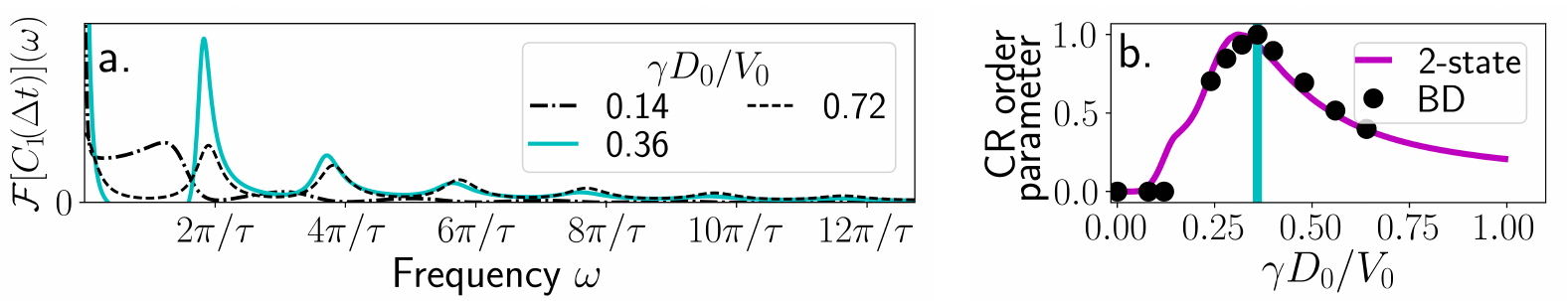}
\caption{ 
{
Detecting coherence resonance (CR) in the bistable system. a. Analytical power spectrum (from 2-state model) given by the Fourier transform of $C_1(\Delta t)=$ $\left\langle X(t)X(t-\Delta t) \right\rangle_\mathrm{ss}$ [Eq.~(\ref{EQ:C1_Tsimring})] at three thermal energies: at and around the value related to CR (0.36). b. Order parameter: Normalized main peak at $\omega \approx 2\pi/\tau$ of the power spectrum of $C_1$, from BD and from the 2-state model. The vertical line at $0.36 \gamma D_0/V_0$ indicates the maximum around which the system is coherence-resonant.
}
}\label{FIG4}
%
%
%
%
\end{figure}

The numerical data in Fig.~\ref{FIG3} reveals that, when the thermal energy is small compared to $V_0$, $\dot{Q}$ increases linearly with $\gamma D_0$, as it also does for a linear delayed system [see Eq.~(\ref{EQ:W4})]. In this regime, approximating the doublewell by its linearised version (dashed line) even renders the correct slope. 
By contrast, in the 2-state model $\dot{Q}=0$, as expected, since the intrawell dynamics is neglected. At larger $\gamma D_0/V_0$, the slope of $\dot{Q}$ abruptly changes and nonlinear behaviour sets in. This occurs at the onset of the delay-induced oscillations, as reflected by {a sudden} increase of the CR order parameter from the 2-state model~\cite{Tsimring2001} {and as confirmed by numerical simulations (see Fig.~\ref{FIG4}). The order parameter measures} the height of the main peak {at a frequency of about $2\pi/\tau$} in the power spectrum of $C_1(\Delta t)$. Hence, a nonzero order parameter indicates the occurrence of subsequent escape events with period $\tau$.

{By further increasing $\gamma D_0/V_0$, one enters the regime where delay-induced oscillations dominate the dynamics. At a certain finite thermal energy (of about $0.36 \gamma D_0/V_0$ in Fig.~\ref{FIG3} and \ref{FIG4}, see cyan vertical lines and cyan plot in Fig.~\ref{FIG4}a.), the peak in the power spectrum is highest resulting in a maximum of the order parameter and indicating that the delay-induced oscillations with mean period $\approx \tau$ are most pronounced. In this range of thermal energies, the escape times are comparable with the delay time $\tau$, and the pronounced peak in the power spectrum signals the resonant response of the system to the noise. Please note that the escape rates can be estimated by the Kramers rates in the corresponding system without delay ($\tau=0$) (see \textit{Technical aspects} section).} {We find that in this range of $\gamma D_0/V_0$}, the 2-state reduction renders a good approximation. In particular, it predicts accurately that a region of steep slope is followed by a lower slope of $\dot{Q}$ accompanied by a maximum of $\dot{S}_\mathrm{m}$, and that the latter lies in the regime of CR. This is our second main result.
%
%
%
%
%
\subsubsection*{{Analytical} explanation for the maximum of entropy production at CR}
The behaviour {of $\dot{Q}$ and $\dot{S}_\mathrm{m}$} can be understood on the basis of the 2-state model. The delay-induced oscillations set in and pause randomly. They have mean period $\tau$, such that $C_1(\tau)=1$ for a perfect oscillation (as in the case of no jumps). However, due to their stochastic nature, $C_1(\tau)$ is lowered with the occurrence of oscillatory events. When the timescales of noise{-induced escapes} and oscillatory motion become comparable, the particle dynamics responds resonantly to the noise. Slightly increasing $\gamma D_0/V_0$ then significantly increases the number of occurring oscillation periods, yielding a strong reduction of $C_1(\tau)$. $\dot{Q}$ thus steeply increases and $\dot{S}_\mathrm{m}$ reaches high values [Eq.~(\ref{EQ:Heat_2state})]. At the CR maximum (vertical {cyan} lines in Fig.~\ref{FIG3}{,\ref{FIG4}}), this effect saturates resulting in a reduced slope of $\dot{Q}$ and a maximum of $\dot{S}_\mathrm{m}$ at CR.

For even higher $\gamma D_0$, the superimposing noise generates irregular, low correlated motion with $C_1(\tau)\rightarrow 0$, hence $\dot{S}_\mathrm{m} \rightarrow 0$. 
In contrast, $\dot{S}_\mathrm{m}$ of the full system approaches a (nonzero) constant (not shown here). The breakdown of the 2-state approximation in this limit is indeed expected since the state discretisation then becomes meaningless. We have tested several parameter settings confirming that the $\dot{S}_\mathrm{m}$ is indeed maximal at CR conditions as predicted by the 2-state model. Quantitatively, we always observe an overestimation of the thermodynamic quantities by a factor of about three. This is somewhat surprising since the phase-space reduction inherent to the discretisation is rather expected to yield an \textit{underestimation} of the steady state quantities~\cite{Esposito2012}. A possible explanation lies in the quasistatic approximation, but the precise reason is subject of future investigations.
\section*{Conclusions}
In this work we put forward analytical approaches for the heat rate in a prototypical nonlinear delayed system.
We have shown that the heat rate can be evaluated based on positional moments, implying that, despite the inherent memory, no temporal correlations are needed. Our formula can be combined with established approximations for the one-time PDF. We have presented analytical results for the Markovian limits and thereby {observed a growing heat flow for vanishing delay time. We have further} uncovered discontinuous behaviour at the onset of memory, {which can also be found for underdamped motion when the delay appears in the velocity. This finite heat flow at small $\tau$-values above and around the (ballistic) velocity-relaxation timescale, is a consequence of the interplay of noise and delay, while the discontinuity at $\tau\to 0$ is a consequence of the white noise assumption and the overdamped limit.
}
For experimental realisations of feedback traps this {theoretical result} implies that the (unavoidable) delay causes a finite steady-state heat flow unless the delay is {significantly} smaller than the (ballistic) relaxation timescale.

{Although the here presented investigations are restricted to static energy landscapes, we indeed expect the results to also hold for time-dependent external potentials, at least, if their changes are slow compared to the other dynamics.}
{ 
Feedback loops offer a way to precisely control the effective temperature of a particle by suppressing its thermal fluctuations\cite{Lee2018}. When this technique is used to change temperature, e.\,g., during the cyclic process of a heat engine\cite{Martinez2017}, the here discussed heat flow, which is caused even by tiny delays, might play a non-negligible role and should therefore be taken into account in the theoretical calculations of efficiency and entropy production.}

 Moreover, we have presented an approach capturing jump processes, thereby predicting that the medium entropy production due to the feedback is maximal, when the delay-induced oscillations are coherence-resonant. {This work is an important step towards an understanding of thermodynamic notions in delayed systems.} Future work will focus on the (non-Gaussian) heat distributions $P(q)$, which appear to violate fluctuation relations \cite{Rosinberg2017} and on developing approaches for the \textit{total} entropy production, possibly via Markovian embedding techniques \cite{Puglisi2009}. 
We hope that our current findings will stimulate experimental investigations on passive (or even active) colloidal systems, e.\,g., a validation of the predicted switching from feedback cooling to heating by adjusting the delay force strength, which is tunable in experimental setups.
\section*{Technical aspects}
\subsection*{{Fokker-Planck equation}}
{
The Fokker-Planck equation of the considered non-Markovian system (\ref{EQ:LE}) is an infinite hierarchy of coupled equations \cite{Guillouzic1999, Loos2017, Frank2005a}, whose first
member reads
\begin{align}\label{EQ:FPE}
\gamma\,\partial_t \rho(x,t)  = &-\partial_x \int_{-\infty}^{\infty} \!\!{F(x,x_\tau)} \rho_\mathrm{2}(x_{\tau},t-\tau;x,t)\mathrm{d}x_{\tau}  
+ \gamma D_0 \partial^2_{xx}\rho(x,t) .
\end{align}
Apart from the one-time PDF, $\rho$, Eq.~(\ref{EQ:FPE}) also involves the two-time joint PDF, $\rho_\mathrm{2}$, and is hence not self-sufficient. 
}
\subsection*{{Derivation of relations between correlation functions (\ref{EQ:Moments-FPE},\ref{EQ:Moments-FPE-LE})}}
{In order to derive (\ref{EQ:Moments-FPE}), we project the Fokker-Planck equation~(\ref{EQ:FPE}) onto its moments by multiplying it with $x^{n+1}$, $n\ge 0$, and integrating over the spatial domain $x\in (-\infty,\infty)$. Several partial integrations, {(where the boundary terms vanish)} yield
\begin{align}\label{EQ:Moments-FPE2}
{\gamma}\frac{\mathrm d }{\mathrm d t}\frac{\left\langle  X(t)^{n+1} \right\rangle}{(n+1)}   =& {\gamma}D_0 \,n   \left\langle  X(t)^{n-1} \right\rangle_\mathrm{ } 
- \!\!\sum_{i=1}^{m}  {a_i}  \left\langle  X(t)^{n+i} \right\rangle_\mathrm{ } 
- b \left\langle  X(t)^{n} X(t-\tau) \right\rangle_\mathrm{ }.
\end{align} 
In a NESS, the left side vanishes and (\ref{EQ:Moments-FPE2}) reduces to a relation between positional moments and the spatial autocorrelation function, as given in (\ref{EQ:Moments-FPE}).
}
{
On the other hand, one can deduce directly by plugging in the LE~(\ref{EQ:LE})
\begin{align}\label{EQ:Moments-LE}
&\frac{\mathrm d }{\mathrm d t}\frac{\left\langle  X(t)^{n+1} \right\rangle}{(n+1)}   = \left\langle \! {X(t)^{n}} \frac{\mathrm{d} X(t)}{\mathrm{d}t }\!\right\rangle \stackrel{\mathrm{LE}}{=}-\!\!\sum_{i=1}^{m} \frac{a_i}{\gamma} \left\langle  X(t)^{i+n} \right\rangle  
+\sqrt{2D_0} \left\langle  X(t)^{n} \xi(t) \right\rangle-\frac{b_{}}{\gamma} \left\langle  X(t)^{n}  X(t-\tau) \right\rangle.
\end{align}
A comparison of Eqs.\,(\ref{EQ:Moments-LE}) and (\ref{EQ:Moments-FPE2}) readily provides the surprisingly simple and generic relation (\ref{EQ:Moments-FPE-LE}), which, in fact, generally holds for any (nonlinear) force (as can be shown analogously).
}
\subsection*{{Derivation of noise--position (\ref{EQ:Contin-x-dx-Corr-underdamped}) and noise--velocity cross correlations for underdamped dynamics}}
{
We aim to calculate the correlations $\langle X(t-\tau)\xi(t)\rangle$ and $\langle \dot{X}(t-\tau)\xi(t)\rangle$ for underdamped motion with position-dependent feedback. To this end, we first we use again the causality argument\cite{Kheifets2014} that the noise cannot influence past position or past velocity. This implies that both correlations must vanish at $\tau>0$. Second, we rewrite the correlations at $\tau=0$ as follows, starting from a formal integration over the LE~(\ref{EQ:LE-underdamped})
\begin{align}
m  \dot{X}(t) =& - \gamma  {X}(t) +  \sum_{i}a_i \int_{0}^{t}X(s)^i\mathrm{d}s -b \int_{0}^{t} X(s-\tau) \mathrm{d}s  + \gamma\sqrt{2 D_0}\int_{0}^{t} \, \xi(t)\mathrm{d}s +\mathcal{C}
\\
\Rightarrow m \langle \dot{X}(t) \xi(t) \rangle = & - \gamma \left\langle   {X}(t)\xi(t) \right\rangle + \sum_{i}\alpha_i \int_{0}^{t} \left\langle X(s)^i\xi(t)\right\rangle \mathrm{d}s -b \int_{0}^{t}\left\langle X(s) \xi(t)\right\rangle \mathrm{d}s  + \gamma\sqrt{ 2D_0} \int_{0}^{t} \left\langle \xi(s)\xi(t)\right\rangle\mathrm{d}s\nonumber \\
\Leftrightarrow m \langle \dot{X}(t) \xi(t) \rangle = &  - \gamma \left\langle  {X}(t)\xi(t) \right\rangle + \sum_{i} a_i \int_{0}^{t} \left\langle X(t)^i\xi(t)  \right\rangle\delta_{s,t} \mathrm{d}s -b \int_{0}^{t}\left\langle X(t) \xi(t)\right\rangle  \delta_{s,t}\mathrm{d}s + \gamma\sqrt{D_0/2}, \label{EQ:Corr_Xdot-Xi}
\end{align} %
where we have simplified the
integrals of types $\int_{0}^{t}\left\langle {X}(s)^i\xi(t)\right\rangle\mathrm{d}s$ by using 
the causality argument which implies that the integrands are zero for all $s<t$, where the constant $\mathcal{C}$ amounts for the initial conditions.%
Third, using the identity ${X}(t) = \int_{0}^{t} \dot{X} (s) \mathrm{d}s+{X}(0)$, we find
\begin{align}\label{EQ:Corr_X-Xi}
\left\langle X(t) \xi(t) \right\rangle = \int_{0}^{t}\left\langle \dot{X}(s) \xi(t) \right\rangle\mathrm{d}s = \int_{0}^{t} \langle \dot{X}(t) \xi(t) \rangle\delta_{s,t}\,\mathrm{d}s .
\end{align} %
Fourth, we establish that Eqs.~(\ref{EQ:Corr_Xdot-Xi},\ref{EQ:Corr_X-Xi}) only allows for finite solutions. This can be shown by contradiction, starting with the assumption $\langle \dot{X}(t) \xi(t) \rangle \to \infty$. Eq.~(\ref{EQ:Corr_X-Xi}) then yields a finite value of $\left\langle X(t) \xi(t) \right\rangle$, which, on the other hand, yields a finite value of $\langle \dot{X}(t) \xi(t) \rangle$ from Eq.~(\ref{EQ:Corr_X-Xi}), thus a contradiction.
Using this finiteness, we finally obtain for $\tau\geq 0$ 
\begin{align}\label{EQ:Correlations_underdamped}
\left\langle {X}(t-\tau) \xi(t) \right\rangle =&\, 0,\\
\left\langle \dot{X}(t-\tau) \xi(t) \right\rangle =&\,(\gamma/m)\, \sqrt{D_0/2 }\,\delta_{\tau}.
\end{align}}
\subsection*{Numerical simulations}
We perform Brownian dynamics simulations with Euler-Maruyama integration scheme. The temporal discretisation is varied between $\Delta t = 10^{-6}$ and $10^{-4}$, such that the typical time scales are all properly resolved, in particular,
the delay time $\tau > 1000\Delta t$ and the intrawell relaxation time $t_\mathrm{rel} > 1000\Delta t$. Furthermore, steady-state ensemble averages are obtained after appropriate transient times (at least $5\times10^6$ time steps) have been cut off, and by using a sufficiently high number of realisations (at least  $10^4$). Both is checked by insensitivity against further increase of simulated time or sample size. Numerical integrations are performed in Stratonovich calculus.
\subsection*{{Escape times}}
{The mean escape time $\tau_\mathrm{K}$ over a barrier of height $\Delta U$ of a Markovian system with potential $U$ can estimated by the Arrhenius formula~\cite{ Loos2017, Tsimring2001, Kramers1940, Hanggi1990} 
\begin{equation} \label{EQ:Arr}
\tau_\mathrm{K}^{-1}=p(x_\mathrm{min})=(2\pi\gamma)^{-1}\sqrt{|U''(x_\mathrm{max})|U''(x_\mathrm{min})} \exp[-\Delta U/(\gamma D_0)],
\end{equation}
where $p(x_\mathrm{min})$ denotes the escape rate out of $x_\mathrm{min}$. %
When using the corresponding system without delay ($\tau=0$) to estimate the escape times in the controlled bistable system, one finds from (\ref{EQ:Arr}) $\tau_K = 2\pi \gamma x_0^2/\sqrt{32}V_0 e^{V_0/D_0\gamma}$, which means $\tau_\mathrm{K}=\tau$ at $0.32 \gamma D_0/V_0$. Hence, around the CR peak, the escape times are comparable with the delay time, as confirmed by our BD results.}%
\subsection*{Reduction to 2-state model}
Within the 2-state model, the particle being in one of the two potential wells is represented by one discrete state $s=\pm x_0$. The transition rates $p_{j \in \{1,2\}}$ between the states dependent on the delayed position, with $p_1$ if $s(t)s(t-\tau) > 0$, and $p_2$ if $s(t)s(t-\tau) < 0$. Within a quasistatic approximation (appropriate if $\tau\gg t_\mathrm{rel}$ and $\gamma D_0 <V_0$), we consider the corresponding quasistatic potential $V_\mathrm{qs}(x) = V_\mathrm{s}(x)+V_\mathrm{d}(x,x_{\tau}=\pm x_0)$, obtained by fixing $x_\tau$ at one of the two deterministically stable solutions $\pm x_0$. The transition rates can then be approximated by the two Kramers rates associated with $V_\mathrm{qs}$, given by {(\ref{EQ:Arr})} {with $p_j=p(x_\mathrm{min,j})$} for escapes out of the two minima, $x_\mathrm{min,1}=\pm x_0$ and $x_\mathrm{min,2}=\mp (x_0/2)[1+\sqrt{1-(k/V_0)}] \approx \mp x_0$, over the barrier at $x_\mathrm{max}=\mp (x_0/2)[1-\sqrt{1-(k/V_0)}]\approx 0$.
In the last step we have used twice $ \sqrt{1-(k/V_0)} \approx 1$ for $k\ll V_0$.
The corresponding second derivatives can be calculated exactly as $V_\mathrm{qs}''(x_\mathrm{max}) = -(4V_0-k)/x_0^2$ and $V_\mathrm{qs}''(x_\mathrm{min,j\in\{1,2\}})  = (8V_0+k)/x_0^2$, while the potential barrier heights read $\Delta V_\mathrm{qs,1} = V_0+{k}/{2},$ and $\Delta V_\mathrm{qs,2}  = V_0-{3k}/{2}$. This finally yields the transition rates
\begin{align}
p_{j \in \{1,2\}}=  {\sqrt{(8V_0+k)(4V_0-k)}}/ \left({2\pi \gamma \,x_0^2} \right) \, \exp \left(- {\Delta V_\mathrm{qs,j}}/{\gamma D_0}\right),
\end{align}
which are the only ingredients needed to calculate $C_1(\Delta t)$ 
from the formula\cite{Tsimring2001}
{ 
\begin{align}\label{EQ:C1_Tsimring}
C^1(\Delta t)=\frac{(\sqrt{p_1}+\sqrt{p_2})\,e^{-2\sqrt{p_1 p_2}\,\Delta t}+(\sqrt{p_1}-\sqrt{p_2})\,e^{-2\sqrt{p_1 p_2}\,( \tau-\Delta t )}}{\sqrt{p_1}+\sqrt{p_2}+(\sqrt{p_1}-\sqrt{p_2})\,e^{-2\sqrt{p_1 p_2}\,\tau}}.
\end{align}
}
\section*{Data Availability}
The numerical data generated and analysed during this study are available from the corresponding author on reasonable request.
\bibliography{AA_references.bib}
\section*{Acknowledgements}
Funded by the Deutsche Forschungsgemeinschaft (DFG, German Research Foundation) - Projektnummer 163436311 - SFB 910. We further thank R. Klages, G. Schaller, and P. H\"ovel for fruitful discussions.
\section*{Author contributions statement}
Both authors formulated the mathematical model. S.A.M.L performed the analysis and simulations. Both authors
interpreted results. S.A.M.L wrote the manuscript with contributions from S.H.L.K. Both authors read and
approved the final manuscript.
\section*{Additional information}
\textbf{Competing interests:} 
The authors declare that they have no competing interests.
\end{document}